\documentclass[aps,preprint,twocolumn,superscriptaddress,10pt]{revtex4}

\usepackage{mciteplus}
\usepackage{amsfonts}
\usepackage{amsmath}
\usepackage{amssymb}
\usepackage{mathtools}
\usepackage{eurosym}
\usepackage{graphicx}
\usepackage{dcolumn}
\usepackage{color}
\usepackage{euscript}
\usepackage{multirow}

\providecommand{\U}[1]{\protect\rule{.1in}{.1in}}
\def\etal{$\textit{et al.}$}
\def\abinitio{$\emph{ab initio}$}
\newcommand{\rom}[1]{\uppercase\expandafter{\romannumeral #1\relax}}

\newcommand{\redbf}[1]{\textcolor{red}{}}

\newcommand{\uu}{$\uparrow\uparrow$}
\newcommand{\ud}{$\uparrow\downarrow$}
\newcommand{\du}{$\downarrow\uparrow$}
\newcommand{\dd}{$\downarrow\downarrow$}
\newcommand{\lrarrow}{$\leftrightarrow$}
\newcommand{\pmau}{$|\pm1,\uparrow\rangle$}
\newcommand{\pmad}{$|\pm1,\downarrow\rangle$}
\newcommand{\pmbu}{$|\pm2,\uparrow\rangle$}
\newcommand{\pmbd}{$|\pm2,\downarrow\rangle$}

\newcommand{\mpad}{$|\mp1,\downarrow\rangle$}

\newcommand{\mpbd}{$|\mp2,\downarrow\rangle$}

\newcommand{\req}[1]{Eq.~(\ref{#1})}
\newcommand{\rsec}[1]{Sec.~\ref{#1}}
\newcommand{\rfig}[1]{Fig.~\ref{#1}}
\newcommand{\rtbl}[1]{Table~\ref{#1}}

\begin{document}
\title{Band-filling effect on magnetic anisotropy using a Green's function method}

\begin{abstract}
We use an analytical model to describe the magnetocrystalline anisotropy energy (MAE) in solids as a function of
band filling. The MAE is evaluated in second-order perturbation theory, which makes it possible to decompose the
MAE into a sum of transitions between occupied and unoccupied pairs.  The model enables us to characterize the MAE
as a sum of contributions from different, often competing terms.  The nitridometalates
Li$_{2}$[(Li$_{1-x}$T$_{x}$)N], with $T$=Mn, Fe, Co, Ni, provide a system where the model is very effective because
atomic like orbital characters are preserved and the decomposition is fairly clean.  Model results are also
compared against MAE evaluated directly from first-principles calculations for this system.  Good qualitative
agreement is found.
\end{abstract}

\author{Liqin Ke}
\email[Corresponding author: ]{liqinke@ameslab.gov}
\affiliation{Ames Laboratory U.S. Department of Energy, Ames, Iowa 50011, USA}
\author{Mark van Schilfgaarde}
\affiliation{Department of Physics, King's College London, Strand, London WC2R 2LS, United Kingdom}
\date{\today}
\maketitle

\section{Introduction}

\label{sec:1}

Magnetocrystalline anisotropy is a particularly important intrinsic magnetic property\cite{rau.science2014}.
Materials with perpendicular magnetic anisotropy are used in an enormous variety of applications, including
permanent magnets, magnetic random access memory, magnetic storage devices, and other spintronics
applications.\cite{mccallum.arms2014,cirera.che2009,rinehart.cs2011,gomezcoca.jacs2013}

Modern band theory methods have been widely used to investigate the magnetocrystalline anisotropy energy (MAE) in
many systems\cite{ravindran.prb2001,solovyev.prb1995}. The MAE in a uniaxial system can be obtained by calculating
the total-energy difference between different spin orientations (out of plane and in plane). However, MAE is
usually a small quantity and a reliable {\abinitio} calculation requires very precise, extensive calculations.
Moreover, MAE is, in general, harder to interpret from the electronic structure than other properties, such as the
magnetization.  MAE often depends on very delicate details of the electronic
structure\cite{belashchenko.apl2015}. Using perturbation theory, the MAE can be decomposed into virtual transitions
between different orbital pairs. In practice, the $d$ bandwidth is large enough that it is nontrivial to
meaningfully resolve the MAE into orbital components and predict its dependence on band filling.

The magnetocrystalline anisotropy originates from spin-orbit coupling (SOC)\cite{vleck.pr1937} or, more precisely,
the change in SOC as the spin-quantization axis rotates.  Including the relativistic corrections to the Hamiltonian
lowers the system energy and breaks the rotational invariance with respect to the spin-quantization axis. Here we
refer to the additional energy due to the relativistic correction as SOC energy or relativistic energy $E^r$. MAE
is a result of the interplay between SOC and the crystal field\cite{larson.prb2004}. The MAE and change in orbital
moment on rotation of the spin-quantization axis are closely related.  We describe this below and denote them as
$K$ and $K_L$, respectively. Without the SOC, the orbital moment is totally quenched by the crystal field in
solids. Except for very heavy elements such as the actinides, SOC usually alleviates only a small part of the
quenching and induces a small orbital moment relative to the spin moment.  For $3d$ transition metals, SOC is often
much smaller than the bandwidth and crystal field splitting, and thus can be neglected in a first
approximation. While the $E^r$ is generally small, its anisotropy with respect to spin rotation is often even
orders of magnitude smaller.

Recently, it had been found that a very high magnetic anisotropy can be obtained in $3d$ systems such as lithium
nitridoferrate
Li$_{2}$[(Li$_{1-x}$Fe$_{x}$)N]\cite{klatyk.prl2002,novak.prb2002,antropov.prb2014,antonov.baps2013}, which can be
viewed as an $\alpha $-Li$_{3}$N crystal with Fe impurities. As found both in experiments\cite{jesche.nphy2014} and
calculations\cite{novak.prb2002,antropov.prb2014} using density functional theory (DFT), the
Li$_{2}$(Li$_{1-x}$Fe$_{x}$)N system possesses an extraordinary uniaxial anisotropy that originates from Fe
impurities. The linear geometry of Fe-impurity sites results in an atomic like orbital and then a large MAE.  As
found in both x-ray absorption spectroscopy\cite{klatyk.prl2002} and DFT
calculations\cite{klatyk.prl2002,novak.prb2002,antropov.prb2014}, $3d$ ions $T$ have an unusually low oxidation
state (+1 ) in Li$_{2}$(Li$_{1-x}T_{x}$)N for $T$= Mn, Fe, Co, and Ni. Recently, Jesche
{\etal}\cite{jesche.prb2015} developed a single-crystal growth technique for these systems and directly observed
that the MAE oscillates when progressing from $T$=Mn$\to$Fe$\to$Co$\to$Ni.\cite{jesche.prb2015} Electronic
structure calculations also show that the atomic like orbital features are preserved for different $T$
elements. Considering the rather large MAE and well-separated density of states (DOS) peaks in this system, it
provides us with a unique platform to investigate the MAE as a function of band filling.

Li and N are very light elements with $s$ and $p$ electrons, respectively. They barely contribute to the MAE in
Li$_{2}$[(Li$_{1-x}T_{x}$ )N]; rather, MAE is dominated by single-ion anisotropy from impurity $T$ atoms,
especially for lower $T$ concentration, where $T$-$T$ atoms become well separated. In this work, we investigate the
magnetic anisotropy with different $T$ elements based on second-order perturbation theory by using a Green's
function method. Lorentzians are used to represent local impurity densities of states and calculate the MAE as a
continuous function of band filling. First-principles calculations of MAE are also performed to compare with our
analytical modeling.

The present paper is organized in the following way. In \rsec{sec:2}, we overview the general formalism of the
single-ion anisotropy\cite{ebert.jap1990,schick.1992} with Green's functions and second-order perturbation
approach\cite{yosida.ptp1965,abate.pr1965,takayama.prb1976,bruno.prb1989,cinal.prb1994,laan.jpcm1998}. Analytical
modeling and calculational details are discussed. In \rsec{sec:3}, we discuss the scalar-relativistic electronic
structure of these systems. The band-filling effect on MAE in Li$_{2}$[(Li$_{1-x}T_{x}$ )N], with $T$=Mn, Fe, Co,
and Ni, is examined within our analytical model and results are compared with first-principles DFT
calculations. The results are summarized in \rsec{sec:4}.

\section{Theory and computational details}
\label{sec:2}
\subsection{Perturbation theory of the magnetocrystalline anisotropy and orbital
moment}

Perturbation theory allows us to calculate magnetic anisotropy directly from the unperturbed band structure.
Orbital moment, SOC energy, and their anisotropies can be written in terms of the
susceptibility.\cite{solovyev.prb1995,ebert.jap1990,takayama.prb1976,cinal.prb1994} The relativistic energy $E^r$
due to the spin-orbit interaction $\Delta V_{so}$=$\xi\mathbf{L}{\cdot}\mathbf{S}$ can be written as

\begin{equation}
E^r=-\frac{1}{2}%
{\int_{-\infty}^{E_{F}}\frac{dE}{\pi}\Im(Tr[\mathbf{G}(E)\Delta V_{so}])}
\label{eqn_k_full}%
\end{equation}

where $\mathbf{G}(E)$ is the full Green's function, which includes SOC and can be constructed from the
non-perturbed Green's function $\mathbf{G}_{0}$. Using second-order perturbation theory (here we consider only
systems with a uniaxial geometry), the relativistic energy can be written as

\begin{equation}
\begin{aligned} 
\label{eqn_k2_ij}%
E^r &= -\frac{1}{2}\Im\sum_{ij}{\int_{-\infty}^{E_{F}}\,\frac{dE}{\pi}Tr\{G_0^{ij}(E)\Delta V_{so}^{j}G_0^{ji}(E)\Delta V_{so}^{i}\}} \\ 
    &= -\frac{1}{2} \sum_{i} \xi_{i}^2 \sum_{\sigma=\pm1} \sum_{m,m'}|\langle m\sigma| \vec{l}\cdot \vec{s} |m' \sigma'\rangle|^2 \chi_{mm'}^{\sigma\sigma'(i)}\\
    &  + \text{intersite terms}
\end{aligned}
\end{equation}

Green's functions are represented in a basis of orthonormalized atomic functions $|i,m,\sigma \rangle$, and $i$
labels atomic sites, $m$ subbands (in cubic harmonics), and $\sigma$ the spin. The local susceptibility
$\chi_{mm'}^{\sigma\sigma'}$, characterizing the transition between two subbands $| m,\sigma \rangle$ and $|
m',\sigma' \rangle$, is defined as
\begin{equation}
\chi_{mm'}^{\sigma\sigma'}(E_F)=\chi_{m'm}^{\sigma'\sigma}(E_F)=\int_{-\infty}^{E_{F}}\frac{dE}{\pi} \Im\{ g_m^{\sigma}g_{m'}^{\sigma'}\},
\label{eqn_xij}%
\end{equation}

where $g_m^{\sigma}$ is the unperturbed on-site Green's function. Because we only consider the on-site contribution
of MAE, only the on-site Green's function or local susceptibility is needed to investigate MAE. We further assume
that on-site Green's functions diagonalize in real harmonic space. The angular dependence and band structure
dependence of relativistic energy $E^r$ are decoupled. In the following, we assume that MAE is dominated by a
particular site $i$, and consider only its contribution.

When the spin-quantization axis is along the $001$ direction, the spin-parallel (longitudinal) components of SO
interaction $l_z$ couple orbitals with the same $|m|$ quantum number ($m$=-$m'$), while the spin-flip (transverse)
ones $l\pm$ couple orbitals with different $|m|$ numbers ($|m|$=$|m|\pm$1). Hereafter, we refer to those two types
of coupling as intra-$|m|$ and inter-$|m|$ types, respectively. According to \req{eqn_k2_ij} and absorbing the site
index $i$, the relativistic energy can be written as

\begin{equation}
E^r_{001}=-\frac{\xi^2}{8}\sum_{\sigma=\pm1}\sum_{m,m'}
\left(
A_{mm'}\chi_{mm'}^{\sigma\sigma}+2B_{mm'}\chi_{mm'}^{-\sigma\sigma}
\right)
\label{eqn_er001}%
\end{equation}

Positive-definite coefficients $A$ and $B$ are just the spin-parallel
and spin-flip parts of the $|L\cdot S|^2$ matrix elements. They can be
written as

\begin{equation}
\label{eqn:amx}
A_{mm'} = m^{2}\delta_{m,-m'}
\end{equation}
\begin{equation}
\label{eqn:bmx}
B_{mm'} = \frac{1}{4}(l(l+1)-m(m\pm1))\delta_{|m|,|m'|\pm1}.
\end{equation}

$A$ and $B$ correspond to intra-$|m|$ and inter-$|m|$ transitions, respectively. An interesting property of the
coefficient matrices is%

\begin{equation}
\sum_{mm'}B_{mm'}=\sum_{mm'}A_{mm'} \label{eqn_aeqb}%
\end{equation}

For an arbitrary spin orientation other than the $001$ direction, one can either obtain the relativistic energy
$E^r$ by rotating $G_0$\cite{solovyev.prb1995} or $V_{so}$\cite{mori.jpsj1969,li.prb1990} in spin subspace. Here we
use the latter approach and the relativistic energy with spin being along the $110$ direction can be written as

\begin{widetext}
\begin{equation}
E^r_{110}=-\frac{\xi^2}{8}\sum_{\sigma=\pm1}\sum_{m,m'}
\left( B_{mm'}\chi_{mm'}^{\sigma\sigma} + (A_{mm'}+B_{mm'})\chi_{mm'}^{-\sigma\sigma} \right)
\label{eqn_er110}%
\end{equation}
\end{widetext}

Notice that spin-parallel coefficients in \req{eqn_er110} are exactly half of the spin-flip coefficients in
\req{eqn_er001}.  If the susceptibility matrix $\mathbf{\chi}$ is relatively homogeneous with respect to spin, then
according to Eqs.~(\ref{eqn_er001}), (\ref{eqn_aeqb}), and (\ref{eqn_er110}), we should expect the spin-flip
components of the relativistic energy $E^r$ to be about twice as large as the spin-parallel
components\cite{antropov.ssc2014}. This is true for the weakly magnetic atoms in different compounds.

Let us define the orbital moment anisotropy (OMA) and MAE, respectively, as
$K_{L}=\langle\mathbf{L}_{z}\rangle_{001}-\langle\mathbf{L}_{z}\rangle_{110}$ and $K=E^{r}_{110}-E^{r}_{001}$. In
this definition, a positive $K$ indicates that the system has a uniaxial anisotropy. If $K_{L}$ is also positive,
then the system has a larger orbital magnetic moment along the easy axis. Using \req{eqn_er001} and
\req{eqn_er110}, the MAE $K$ can be written as
\begin{widetext}
\begin{equation}
K=\frac{\xi^{2}}{8}\sum_{m,m'}(A_{mm'}-B_{mm'})(\chi_{mm'}^{\uparrow\uparrow}+\chi_{mm'}^{\downarrow\downarrow}-\chi_{mm'}^{\uparrow\downarrow}-\chi_{mm'}^{\downarrow\uparrow}).
\label{eqn_ke}%
\end{equation}
\end{widetext}

\begin{figure}[ht]
\begin{tabular}{c}
\includegraphics[width=.45\textwidth,clip,angle=0]{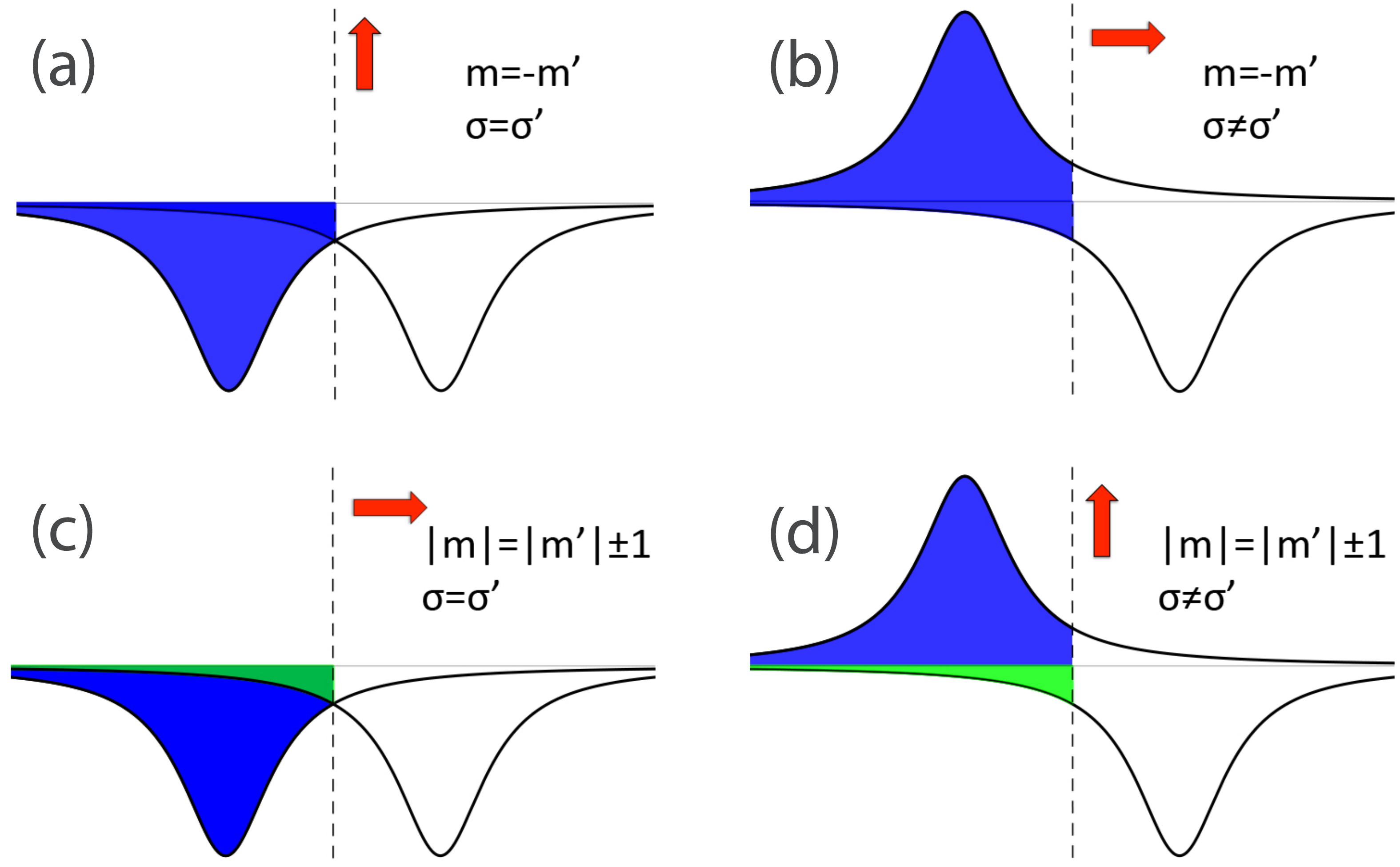}%
\end{tabular}%
\caption{ (Color online) Illustration of the dependence of the easy-axis direction on the orbital quantum numbers
  $(m,m')$ and the spin quantum numbers $(\sigma,\sigma')$ of two subbands. Configurations (a) and (d) favor
  uniaxial anisotropy, while (b) and (c) favor easy-plane anisotropy. The vertical dotted line corresponds to the
  Fermi energy, $E_{F}$. The horizontal line separates the majority (up) and minority (down) spin
  channels. Occupied states with different $|m|$ numbers are filled with different colors.}
\label{fig:kmmss}
\end{figure}

MAE is resolved into allowed transitions between all pairs of orbitals
$|m,\sigma\rangle$\lrarrow$|m',\sigma\rangle$, corresponding to the $\chi_{mm'}^{\sigma\sigma'}$ terms. Since $A$
and $B$ are positive definite, the coefficient of $\chi_{mm'}^{\sigma\sigma'}$ is positive when ($m$=$-m'$ and
$\sigma$=$\sigma'$) or ($|m|$=$|m'|\pm1$ and $\sigma$=$-\sigma'$), and is negative when ($m$=$-m'$ and
$\sigma$=$-\sigma'$) or ($|m|$=$|m'|\pm1$ and $\sigma$=$\sigma'$). In general, the local susceptibility
$\chi_{mm'}^{\sigma\sigma'}$ is also positive definite; hence we have the following simple selection rule for MAE:
For intra-$|m|$ orbital pairs, transitions between same (different) spin channels promote easy-axis (easy-plane)
anisotropy; for inter-$|m|$ pairs, the sign is the other way around, i.e., transitions between same (different)
spin channels promote easy-plane (easy-axis) anisotropy. This simple rule is illustrated in \rfig{fig:kmmss}.

Similarly, the OMA $K_L$ can be written as
\begin{equation}
K_L=\frac{\xi}{2}\sum_{m,m'}(A_{mm'}-B_{mm'})(\chi_{mm'}^{\downarrow\downarrow}-\chi_{mm'}^{\uparrow\uparrow})
\label{eqn_kl}%
\end{equation}

Hence, OMA originates from the difference between $\uparrow\uparrow$ and $\downarrow\downarrow$ components of each
pair susceptibility, while MAE originates from the difference between the spin-parallel and spin-flip
components. If we sum over contributions from all the spin components from each pair of orbitals $(m,m')$ and
define

\begin{equation}
\chi_{mm'}^{\epsilon}=\chi_{mm'}^{\uparrow\uparrow}+\chi_{mm'}^{\downarrow\downarrow}-\chi_{mm'}^{\uparrow\downarrow}-\chi_{mm'}^{\downarrow\uparrow}
\label{eqn_chiep}
\end{equation}

\begin{equation}
\chi_{mm'}^{l}=\chi_{mm'}^{\downarrow\downarrow}-\chi_{mm'}^{\uparrow\uparrow},
\end{equation}

then Eqs.~(\ref{eqn_ke}), and (\ref{eqn_kl}) can be written as

\begin{equation}
\frac{4}{\xi^2}K = \frac{1}{2}\sum_{m,m'}(A_{mm'}-B_{mm'}) \chi_{mm'}^{\epsilon} 
\label{eqn:kexe}
\end{equation}
\begin{equation}
\frac{1}{\xi}K_L = \frac{1}{2}\sum_{m,m'}(A_{mm'}-B_{mm'}) \chi_{mm'}^{l}
\end{equation}

Obviously, the correlation between OMA and MAE\cite{ke.prb2013} only happens when the susceptibility is dominated
only by one of the spin-parallel components. If it is dominated by $\chi^{\uparrow\uparrow}$, then the system has a
smaller orbital moment along the easy axis\cite{antropov.ssc2014}. If it is dominated by
$\chi^{\downarrow\downarrow}$, then the system has a larger orbital moment along the easy axis and we have
$K$=$\frac{\xi}{4} K_{L}$.

Equation (\ref{eqn_ke}) is useful to explain the MAE in two extreme cases.  (i) Nonmagnetic limit: Since the
orbitals are spin independent, we have
$\chi_{mm'}^{\uparrow\uparrow}=\chi_{mm'}^{\uparrow\downarrow}=\chi_{mm'}^{\downarrow\uparrow}=\chi_{mm'}^{\downarrow\downarrow}$.
$\chi_{mm'}^{\epsilon}$ vanishes for every pair of subbands $mm'$ because the spin-parallel components cancel out
the spin-flip ones. (ii) Zero crystal-field limit: Since orbitals are degenerate,
$\sum_{mm'}(A_{mm'}-B_{mm'})\chi_{mm'}^{\sigma\sigma'}$ in \req{eqn_ke} vanishes for each of the four spin
components $\sigma\sigma'$. Thus the total anisotropy vanishes as in a free atom.

Using the expressions of coefficients in Eqs.~(\ref{eqn:amx}) and (\ref{eqn:bmx}), for a $d$-orbital system,
\req{eqn_ke} can be written as
\begin{widetext}
\begin{equation}
\frac{4}{\xi^2}K=4\mathbf{\chi}_{-2,2}^{\epsilon}+\mathbf{\chi}_{-1,1}^{\epsilon}
-\frac{3}{2}\left(  
\mathbf{\chi}_{-1,0}^{\epsilon}+\mathbf{\chi}_{0,1}^{\epsilon}
\right)  
-\frac{1}{2}\left(  
\mathbf{\chi}_{-2,-1}^{\epsilon}+\mathbf{\chi}_{-2,1}^{\epsilon}
+\mathbf{\chi}_{-1,2}^{\epsilon}+\mathbf{\chi}_{1,2}^{\epsilon}
\right)  
\label{eq:k8}%
\end{equation}
\end{widetext}

where the ordering of the states is $|$-2$\rangle$=$d_{xy}$, $|$-1$\rangle$=$d_{yz}$, $|0\rangle$=$d_{z^{2}}$,
$|1\rangle$=$d_{xz}$, and $|2\rangle$=$d_{x^{2}-z^{2}}$.  Different point-group symmetry results in different
orbital degeneracy on site $i$. By summing up the coefficients of equivalent orbital pairs, \req{eq:k8} can be
simplified.

For tetragonal, square planar, or square pyramidal geometries, one pair of orbitals ($d_{xz},d_{yz}$) is
degenerate. Equation (\ref{eq:k8}) can be written as
\begin{equation}
\frac{4}{\xi^2}K= 4\mathbf{\chi}_{-22}^{\epsilon}+\mathbf{\chi}_{11}^{\epsilon
}-\mathbf{\chi}_{12}^{\epsilon}-3\mathbf{\chi}_{01}^{\epsilon}-\mathbf{\chi
}_{-2,1}^{\epsilon}.  %
\label{eq:k4l}%
\end{equation}
We recover Eq.~(13) in Ref.[\onlinecite{takayama.prb1976}].
%% 3 levels

For linear, trigonal, petagonal bipyramidal, and square antiprismatic
geometries, besides ($d_{xz},d_{yz}$) orbitals, ($d_{x^{2}-y^{2}}$
,$d_{xy}$) orbitals are also degenerate. Equation (\ref{eq:k4l}) can
be further simplified as
\begin{equation}
\frac{4}{\xi^2}K=4\mathbf{\chi}_{22}^{\epsilon}+\mathbf{\chi}_{11}^{\epsilon
}-3\mathbf{\chi}_{01}^{\epsilon}-2\mathbf{\chi}_{12}^{\epsilon}%
\label{eq:k3l}%
\end{equation}

%% 2 levels
On the other hand, for tetrahedral and octahedral geometries, five $d$
orbitals split into two groups $E_g$ and $T_{2g}$, namely,
($d_{z^{2}}$, $d_{x^{2}-y^{2}}$) and ($d_{xy}$, $d_{yz}$,
$d_{xz}$). One can easily show that the right side of \req{eq:k8}
vanishes as expected for cubic geometry.

Similarly, with the coefficient matrices and orbital degeneracy, one
easily recovers the formulas for the orbital moment in the tetragonal
system as in Ref.[\onlinecite{ebert.jap1990}] or $A1$ and $A2$ as in
Ref.[\onlinecite{solovyev.prb1995}].

\subsection{band-filling effect on MAE in a two-level model}

\begin{figure}[ptb]%
\begin{tabular}
[c]{c}%
\includegraphics[width=.45\textwidth,clip,angle=0]{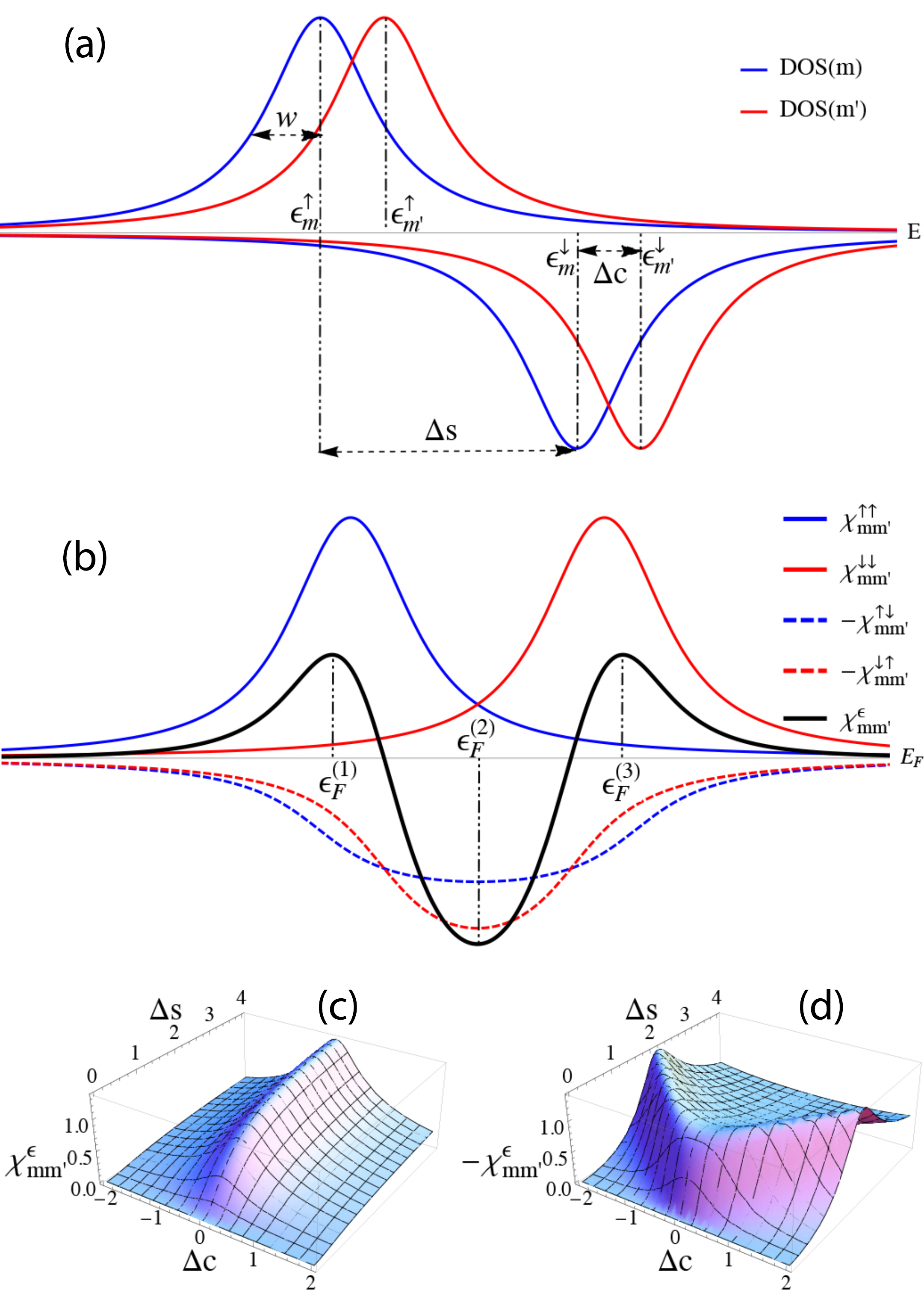}  \\
\end{tabular}
\caption{(Color online) (a) Schematic Lorentzian-shape densities of states for subbands $m$ and $m'$.  (b)
  $\chi_{mm'}^{\epsilon}$ and its four spin components as functions of Fermi energy.  The amplitudes of
  $\chi_{mm'}^{\epsilon}$ with (c) the maximum at $\varepsilon_{F}^{(1,3)}$ and (d) the minimum at
  $\varepsilon_{F}^{(2)}$ as functions of spin splitting $\Delta s$ and crystal-field splitting $\Delta c$.}
\label{fig:xmm}%
\end{figure}

As shown in \req{eqn_ke}, the MAE and OMA can be resolved into contributions from allowed transitions between all
pairs of orbitals. The sign and weight of the contribution are determined by coefficients $A_{m,m'}$ and
$B_{m,m'}$, which only depend on the orbital characters of the corresponding orbital pairs. On the other hand,
$\chi_{mm'}^{\epsilon}$, or its four components $\chi_{mm'}^{\sigma\sigma'}$, are determined by the electronic
structure, namely, the Fermi level (electron occupancy or band filling), band width, crystal-field splitting, and
spin splitting. Here we investigate the band-filling effect on the MAE contribution from a single pair of orbitals.
For each orbital pair $mm'$, there are four spin components: two spin-parallel ({\uu} and {\dd}) terms and two
spin-flip terms ({\ud} and {\du}). As assumed in the Anderson model, Lorentzians are used to represent the local
densities of state (LDOS) in our analytical model to illustrate the electronic structure dependence of
$\chi_{mm'}^{\sigma\sigma'}$ and MAE. Similarly, Ebert {\etal}\cite{ebert.jap1990} used Lorentzians DOS to
analytically investigate the orbital magnetic moment and relate it to the impurity density of states at the Fermi
level. For simplicity, we use the same width for every Lorentzian orbital, and the on-site Green's function for
subband $|m\rangle$ in one spin channel $\sigma$ is given by
\begin{equation}
g_{m}^{\sigma}(E)=\frac{1}{E-\varepsilon_{m}^{\sigma}+iw}%
\end{equation}
where $\varepsilon_{m}^{\sigma}$ is the band center and $w$ is the half width.  The corresponding LDOS for subbands
$|m\rangle$ and $|m'\rangle$ in two spin channels are shown in \rfig{fig:xmm}(a). For simplicity, we further assume
that the two subbands have the same spin splitting,
$\varepsilon_{m}^{\sigma}-\varepsilon_{m}^{\sigma'}$=$\varepsilon_{m'}^{\sigma}-\varepsilon_{m'}^{\sigma'}\equiv\Delta
s$, or, equivalently, have the same crystal-field splitting,
$\varepsilon_{m}^{\sigma}-\varepsilon_{m'}^{\sigma}$=$\varepsilon_{m}^{\sigma'}-\varepsilon_{m'}^{\sigma'} \equiv
\Delta c$, in the two spin channels.

According to \req{eqn_xij}, the pairwise local susceptibility for orbitals $|m,\sigma\rangle$ and $|m',\sigma'\rangle$ can be written as%
\begin{widetext}
\begin{equation}
\chi_{mm'}^{\sigma\sigma^{\prime}}\left(  E_{F}\right)  =\left\{
\begin{array}
[c]{ccc}%
\frac{1}{\pi}\frac{1}{\varepsilon_{m'}^{\sigma
^{\prime}}-\varepsilon_{m}^{\sigma}}(\arctan[\frac{E_{F}-\varepsilon_{m}^{\sigma}}%
{w}]-\arctan[\frac{E_{F}-\varepsilon_{m'}^{\sigma^{\prime}}}{w}]) &
& \text{if }\varepsilon_{m}^{\sigma}\neq\varepsilon_{m'}^{\sigma^{\prime}}\\
D(E_{F})=\frac{1}{\pi}\frac{w}{(E_{F}-\varepsilon_{m}^{\sigma})^{2}+w^{2}} &
& \text{if }\varepsilon_{m}^{\sigma}=\varepsilon_{m'}^{\sigma^{\prime}}%
\end{array}
\right.
\label{eqn_chimmss}
\end{equation}
\end{widetext}

$\chi_{mm'}^{\sigma\sigma'}(E_{F})$ is a positive-definite function for any $E_{F}$ and reaches the maximum at
$E_{F}$=$(\varepsilon_{m}^{\sigma}+\varepsilon_{m'}^{\sigma'})/2$.  The maximum value increases as the two band
centers approach each other until becoming degenerate, because the energies required to transfer electrons from
occupied states to the unoccupied states become smaller. Band narrowing increases $\chi_{mm'}^{\sigma\sigma'}$
quickly (nearly $1/w$) until it reaches the atomic limit. When the bandwidth becomes comparable to or smaller than
the SOC constant, SOC can lift the orbital degeneracy and shift two states, i.e., one above and the other below the
Fermi level $E_F$ completely. On the other hand, if the Fermi level sits between two well-separated narrow subbands
and bandwidth is small compared to the distance between the Fermi level and the two band centers,
$w$$\ll$$E_{F}-\varepsilon_{m}^{\sigma}$ and $w$$\ll$$\varepsilon_{m'}^{\sigma'-E_{F}}$, according to
\req{eqn_chimmss}, then $\chi_{mm'}^{\epsilon}$=$1/(\varepsilon_{m'}^{\sigma
  ^{\prime}}$-$\varepsilon_{m}^{\sigma})$ does not depend on the Fermi energy.

Using Eqs.~(\ref{eqn_chiep}) and (\ref{eqn_chimmss}), the dependencies of $\chi_{mm'}^{\epsilon}$ and its four spin
components on the Fermi energy $E_{F}$ are shown in Fig.\ref{fig:xmm}(b). There is one minimum at
$\varepsilon_{F}^{(2)}$ and two maxima at $\varepsilon_{F}^{(1,3)}$, with

\begin{equation}
\varepsilon_{F}^{(i)}  = \frac{\varepsilon_{1}+\varepsilon_{2}+\bigtriangleup s}{2}
  +  \frac{i-2}{2}\sqrt{(\Delta c)^{2}+(\bigtriangleup s)^{2}+4w^{2}}
\label{eq:xeroot}
\end{equation}

The two maximum peaks originate from the two spin-parallel terms $\chi_{mm'}^{\uparrow\uparrow}$ and
$\chi_{mm'}^{\downarrow\downarrow}$, while the minimum originates from the spin-flip terms
$-(\chi_{mm'}^{\uparrow\downarrow}+\chi_{mm'}^{\downarrow\uparrow})$. In \req{eq:xeroot}, each spin component
$\chi_{mm'}^{\sigma\sigma'}$ has its maximum amplitude when the Fermi level is around the middle of the
corresponding two band centers. The two spin-flip components have their maximum values at the same Fermi level
$\varepsilon_{F}^{(2)}$ because we assume that the two orbitals have the same spin splittings. Contributions from
the two spin-flip components become identical when two states $|m\rangle$ and $|m'\rangle$ are degenerate.

As shown in Eqs.~(\ref{eqn_ke}) and (\ref{eqn:kexe}), the MAE coefficients for intra-$|m|$ ($A$) and inter-$|m|$
terms (-$B$) have different signs. To have a large uniaxial anisotropy, the Fermi level should be around the
$\varepsilon _{F}^{(1)}$ or $\varepsilon_{F}^{(3)}$ for intra-$|m|$ orbital pairs and $\varepsilon_{F}^{(2)}$ for
inter-$|m|$ orbital pairs. Two orbitals can accommodate four electrons in two spin channels, and $\varepsilon
_{F}^{(i)}$ roughly corresponds to band filling of one, two, and three electrons with $i$=1, 2, and 3,
respectively.  Figures \ref{fig:xmm}(c) and \ref{fig:xmm}(d) shows the maximum amplitude of
$\chi_{ij}^{\epsilon}$($E_{F}$=$\varepsilon_{F}^{(i)}$) as functions of crystal splitting $\Delta c$ and spin
splitting $\triangle s$. For $E_{F}$=$\varepsilon_{F}^{(1,3)}$, it requires $\triangle c$=0 to align the two
subbands in the same spin channel (two subbands becomes degenerate).  For $E_{F}$=$\varepsilon_{F}^{(2)}$, it
requires $\triangle s$=$\pm\triangle c$ to align the two subbands in different spin channels.

\subsection{Crystal structures}

\begin{figure}[ht]
\begin{tabular}{c}
  \includegraphics[width=.49\textwidth,clip,angle=0]{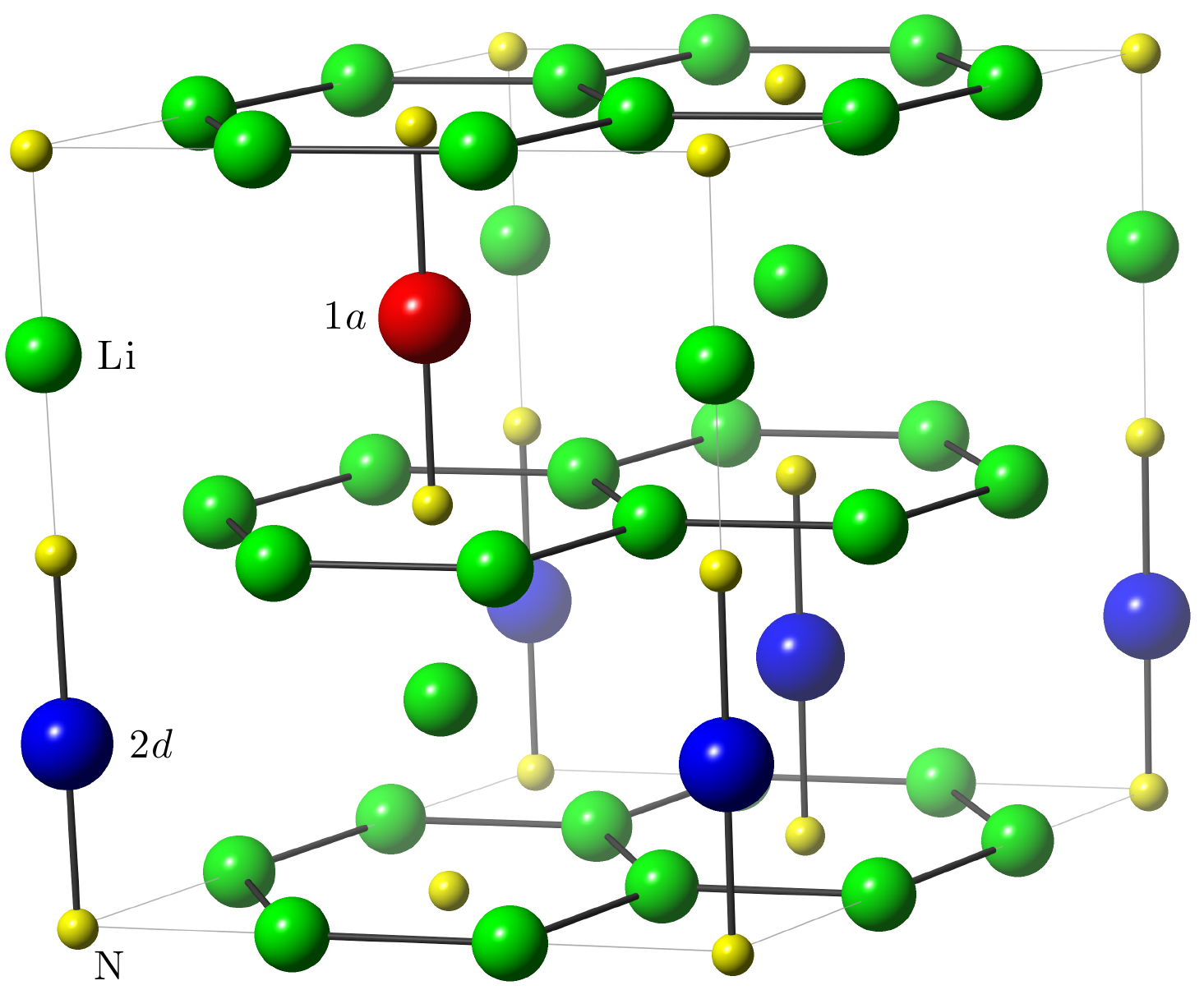}  \\
\end{tabular}%
\caption{ (Color online) Schematic representation of the supercell used in the DFT calculation for
  Li$_{2}$[(Li$_{1-x}$T$_{x}$)N] with $x$=0.5. Both $T_{1a}$ and $T_{2d}$ sites are derived from the Li$_{\rom{1}}$
  ($1b$) site in the original $\alpha$-Li$_{3}$N structure, while other Li atoms, which form coplanar hexagons,
  correspond to Li$_{\rom{2}}$ ($2c$) sites in the original $\alpha$-Li$_{3}$N structure.}
\label{fig:xtal}
\end{figure}

Li$_{2}$(Li$_{1-x}T_{x}$)N crystallizes in the $\alpha$-Li$_{3}$N structure type, which is hexagonal and with space
group $P6/mmm$ (no. 191). The unit cell of $\alpha$-Li$_{3}$N contains one formula unit. There are two
crystallographically inequivalent sets of Li atoms, Li$_{\rom{1}}$ ($1b$) and Li$_{\rom{2}}$ ($2c$), with $6/mmm$
and $-6m2$ point-group symmetries, respectively. The Li$_{\rom{1}}$ atoms are sandwiched between two N atoms and
form a linear -Li$_{\rom{1}}$-N- chain along the axial direction, while Li$_{\rom{2}}$ sites have twofold
multiplicities and form coplanar hexagons which are centered at -Li$_{\rom{1}}$-N- chains and parallel to the basal
plane.  Li$_{\rom{2}}$ is more close packed in lateral directions and $3d$ atoms randomly occupy Li$_{\rom{1}}$
sites. We carried out DFT calculations for small doping concentration with $x$=0.166 and found that all $T$
elements with $T$=Mn, Fe, Co, and Ni indeed prefer to occupy Li$_{\rom{1}}$ sites.  To calculate the electronic
structure and MAE, we use a supercell which corresponds to a $\sqrt3$$\times$$\sqrt3$$\times$2 superstructure of
the original $\alpha$-Li$_{3}$N unit cell. Details of the supercell construction can be found in
Ref.[\onlinecite{novak.prb2002}].  For $x$=0.5, as shown in \rfig{fig:xtal}, there are three $T$ atoms in the
24-atom supercell with one on the $1a$ site and the other two on the $2d$ sites.  Both $T_{1a}$ and $T_{2d}$ sites
are derived from the $1b$ site in the original $\alpha$-Li$_{3}$N. They have a linear geometry and a strong
hybridization with neighboring N atoms along the axial direction. $T_{1a}$ have six Li neighbors, while $T_{2d}$
have three $T_{2d}$ and three Li neighbors in the $T$-Li plane. This structure (denoted as $hex2$ in
Ref.[\onlinecite{novak.prb2002}]) is of particular interest because two types of $T$ sites, $T_{1a}$ and $T_{2d}$,
possess very different local surroundings and represent different local impurity concentrations.  Along the
in-plane direction, $T$-$T$ distances are rather large, especially for the $1a$ site.  Since the $T_{1a}$ site
represents a relatively low impurity concentration and dominates the uniaxial MAE for $T$=Fe, most of the results
in this work are focused on the $T_{1a}$ site in the $hex2$ supercell. We also consider other concentrations such
as $x$=0.16 and $x$=0.33.

\subsection{DFT calculational details}

We carried out first principles DFT calculations using the Vienna {\abinitio} simulation package
(VASP)\cite{kresse.prb1993,kresse.prb1996} and a variant of the full-potential linear muffin-tin orbital (LMTO)
method\cite{methfessel.chap2000}.  We fully relaxed the atomic positions and lattice parameters, while preserving
the symmetry using VASP.  The nuclei and core electrons were described by the projector augmented-wave
potential\cite{kresse.prb1999} and the wave functions of valence electrons were expanded in a plane-wave basis set
with a cutoff energy of 520 eV.  For relaxation, the generalized gradient approximation of Perdew, Burke, and
Ernzerhof was used for the correlation and exchange potentials. The spin-orbit coupling is included using the
second-variation procedure\cite{koelling.jpc1977,shick.prb1997}. We also calculated the MAE by carrying out
all-electron calculations using the full-potential LMTO (FP-LMTO) method to check our calculational results. For
the MAE calculation, the $k$-point integration was performed using a modified tetrahedron method with Bl\"ochl
corrections, with $16^3$ $k$-points in the first Brillouin zone of the 24-atom unit cell. By evaluating the SOC
matrix elements $\langle V_{SO} \rangle$ and its anisotropy\cite{antropov.ssc2014}, we resolve the anisotropy of
orbital moment and MAE into sites, spins, and orbital pairs. The correlation effects are also considered by using
the local-density approximation (LDA)+$U$ method. Here we choose the fully localized limit implementations of the
double counting introduced by Liechtenstein {\etal}\cite{liechtenstein.prb1995} considering it is more appropriate
for materials with electrons localized on specific orbitals.

\section{Results and discussions}
\label{sec:3}

\subsection{Electronic structures}
\begin{figure}[ptb]%
\begin{tabular}
[c]{c}%
\includegraphics[width=.47\textwidth,clip,angle=0]{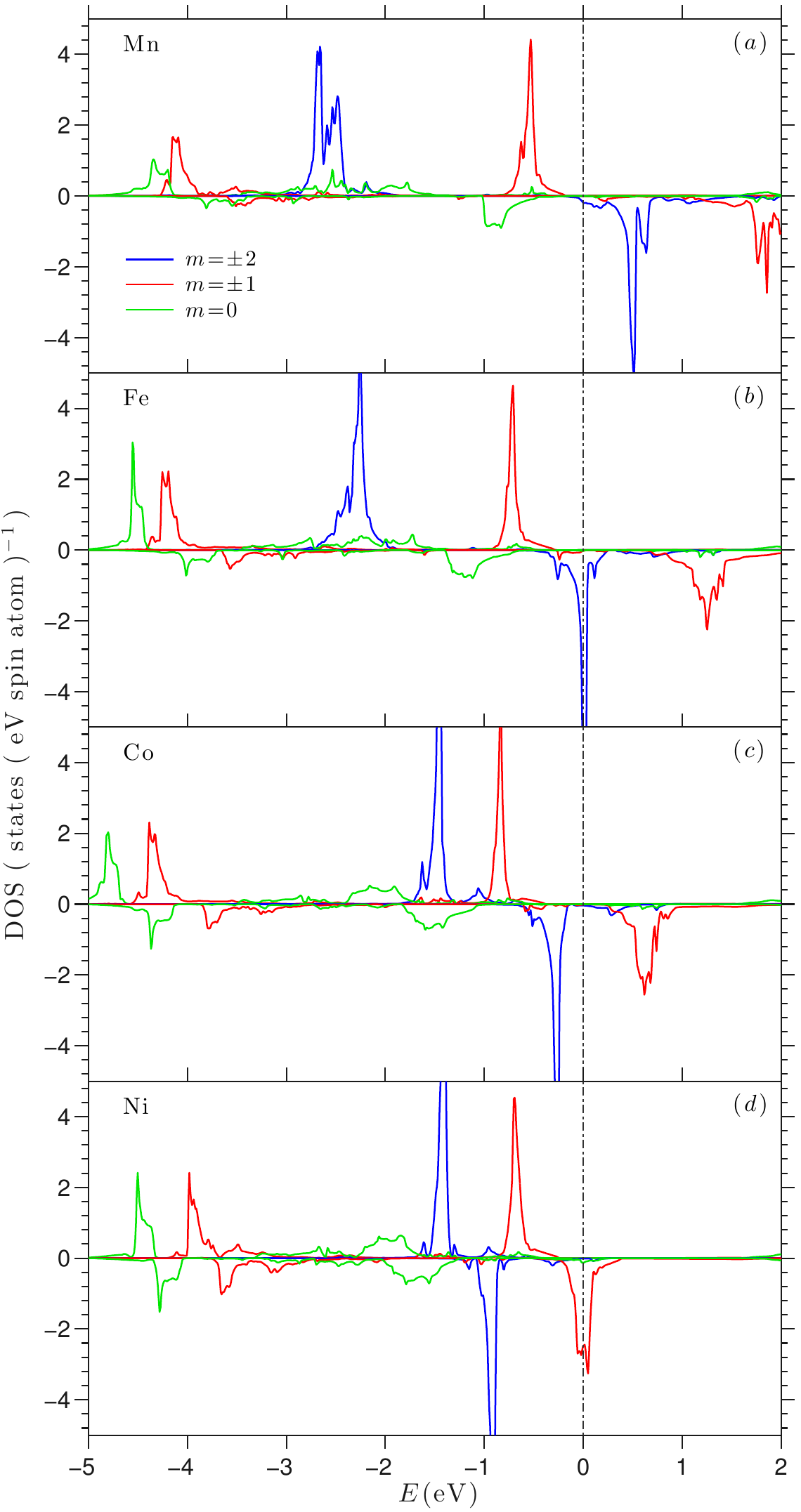}\\
\end{tabular}
\caption{(Color online) Partial densities of states projected on the $3d$ states of the $T_{1a}$ site in the $hex2$
  structure in Li$_{2}$[(Li$_{1-x}$T$_{x}$)N], where $x$=0.5 and $T$ is (a) Mn, (b) Fe, (c) Co, and (d) Ni. The
  vertical dotted line corresponds to the Fermi energy, $E_{F}$. The horizontal dotted line separates the majority
  (up) and minority (down) spin channels. Calculation is within LDA, without spin-orbit coupling included. }%
\label{fig:pdos}%
\end{figure}

Without considering SOC, the axial crystal field on both $T_{1a}$ and $T_{2d}$ sites splits five $3d$ orbitals into
three groups: degenerate ($d_{xy}$, $d_{x^{2}-y^{2}}$) states, degenerate ($d_{yz}$, $d_{xz}$ ) states, and
$d_{z^2}$ state. Equivalently, they can be labeled as $m$=$\pm2$, $m$=$\pm1$, and $m$=$0$ using cubic harmonics.

The scalar-relativistic partial densities of states (PDOS) projected on the $T_{1a}$ site are shown in
\rfig{fig:pdos}.  For $T$=Fe, the PDOS obtained is very similar to what was previously reported
\cite{novak.prb2002}. The Fe $3d$ shell has seven electrons and the majority spin channels of $d$ orbitals are
fully occupied with five electrons. 

The Fe $d_{z^2}$ states hybridize with $p_z$ states of $N$ atoms along the axial direction and mix with on-site
$4s$ states, which causes the $d_{z^2}$ orbital to be lower in energy than the other $d$
orbitals.\cite{novak.prb2002} The $d_{z^2}$ states spread out and lie below the Fermi level and accommodates one
electron in the minority spin channel. The last electron occupies half of the degenerate ($d_{xy}$,
$d_{x^{2}-y^{2}}$) states in the minority spin channel. These states have a very narrow bandwidth and cross the
Fermi level.

The linear geometry minimizes the in-plane hybridization between the $T$ $3d$ orbitals and the neighboring atoms,
making them atomic like and resulting in narrower bands. The $T_{2d}$ site shows a similar PDOS as the $T_{1a}$
site; however, the in-plane hybridization with other $T_{2d}$ sites results in a much broader bandwidth than the
$1a$ sites.

For other $T$ elements, the DOS peaks are well separated as in $T$=Fe. The minority spin channel clearly shows a
different band-filling pattern with different $T$ elements. The deviation from the rigid-band model is also
obvious. Spin splitting decreases from Mn to Ni, while the crystal-field splitting values (the energy difference
between $m$=$\pm1$ and $m$=$\pm2$ states) are larger for $T$=Mn and Fe than for $T$=Co and Ni.

Figure \ref{fig:kmodft}(a) shows the schematic Fe PDOS, and how the Fermi level changes with different $T$ in a
rigid-band approximation (RBA). Different $T$ elements correspond to different integer number of $3d$
electrons. Since each degenerate state pair can accommodate two electrons in one spin channel, the Fermi level
either intersects the degenerate peaks or sits in the middle of two peaks.

\subsection{MAE in Li$_{2}$[(Li$_{1-x}T_{x}$)N]  with $T$=Fe}
%% \label{sec:4}

\begin{table}[ptb]
\caption{Lattice constants, total and site-resolved MAE Li$_{2}$[(Li$_{0.5}T_{0.5}$)N] with $T$=Mn, Fe, Co and
  Ni. The MAE values for $T_{2d}$ site are in unit of $m$eV/atom, and there are two $T_{2d}$ atoms in the supercell.}%
\label{tbl:klitn}
\begin{tabular}
[c]{ccccccc}\hline\hline
& \multicolumn{2}{c}{Lattice parameters} & \multicolumn{4}{c}{$K(meV)$%
}\\\cline{2-3}\cline{4-7}%
\hspace{15pt}T\hspace{15pt} & $a$(a.u.) & $c/a$ & cell & $T_{1a}$ & $T_{2d}$ &
others\\\hline
Mn & 12.143 & 1.202 & -1.14 & -0.35 & -0.38 & -0.03\\
Fe & 12.091 & 1.183 & 20.83 & 14.77 & 3.09 & -0.12\\
Co & 12.144 & 1.154 & -3.69 & -0.89 & -1.32 & -0.15\\
Ni & 12.113 & 1.156 & 2.52 & 1.71 & 0.37 & 0.06\\\hline\hline
\end{tabular}
\end{table}

MAE in Li$_{2}$[(Li$_{1-x}T_{x}$)N] with $T$=Mn, Fe, Co, and Ni and $x=0.5$ are calculated in DFT and summarized in
Table \ref{tbl:klitn}. The system has uniaxial anisotropy with $T$=Fe or Ni and easy-plane anisotropy with $T$=Mn
or Co. MAE is dominated by the contributions from the $1a$ site for $T$=Fe or Ni. Results are in qualitative
agreement with previous calculations.\cite{klatyk.prl2002,novak.prb2002,antropov.prb2014} The extraordinary MAE for
$T=$Fe originates from the unique band structure in this system.  Because the well-isolated Fe atoms, such as the
Fe$_{1a}$ site in the $hex2$ supercell, provide the major contribution to the uniaxial anisotropy, we focus on the
Fe$_{1a}$ site.

\begin{figure}[ptb]%
\begin{tabular}
[c]{c}%
\includegraphics[width=.46\textwidth,clip,angle=0]{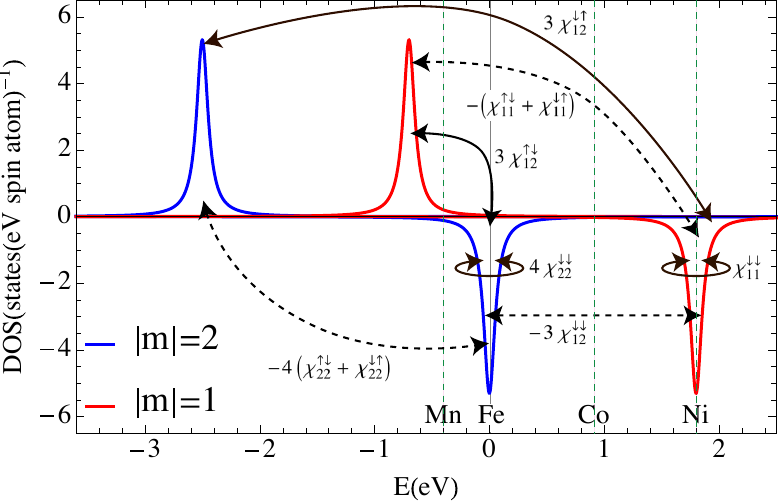}  \\
\includegraphics[width=.46\textwidth,clip,angle=0]{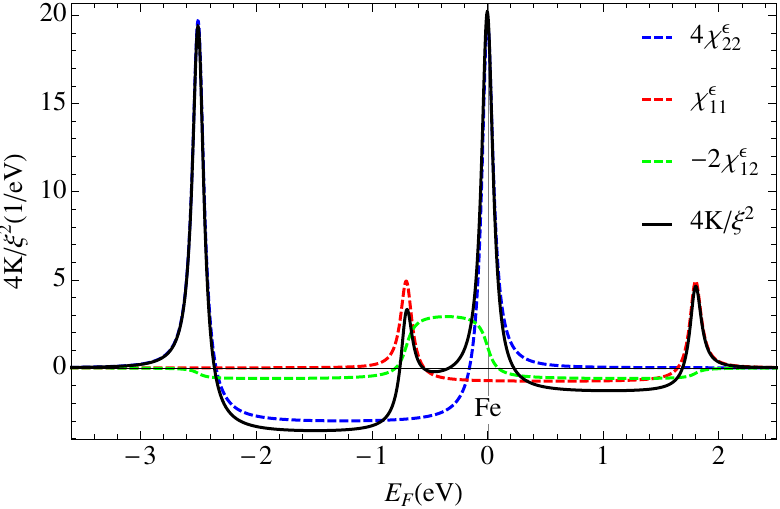} \\
\includegraphics[width=.46\textwidth,clip,angle=0]{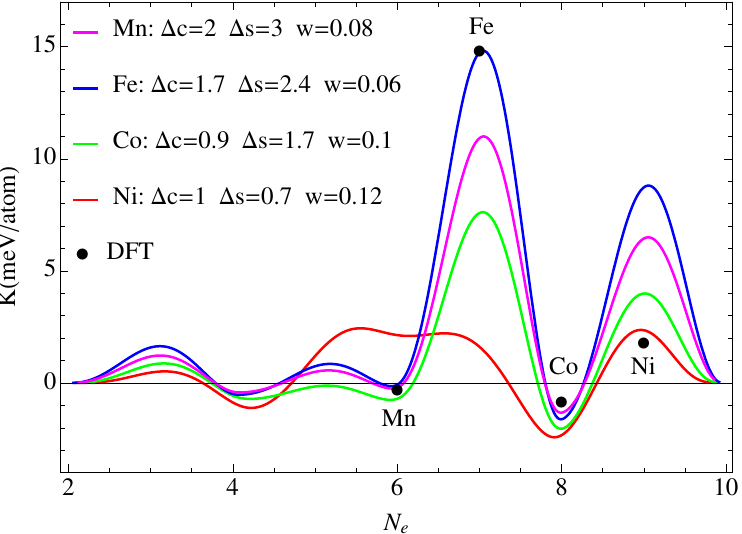} \\
\end{tabular}
\caption{(Color online) (a) Schematic partial densities of states projected on the $3d$ states of Fe$_{1a}$
  sites. Orbital transitions and the sign of their contributions to the MAE are also shown. Solid line indicates
  positive contribution (easy axis) and the dashed line indicates negative contribution (easy plane) to the
  easy-axis anisotropy.  (b) Scaled MAE $4K/\xi^{2}$ from $T_{1a}$ site and its decomposition into orbital
  susceptibilities as functions of band filling. (c) Magnetic anisotropy energy $K$ from $T_{1a}$ site as a function
  of $T$. Different sets of electronic structure parameters $\Delta s$, $\Delta c$, and $w$ are used to represent
  the DFT PDOS on $T_{1a}$ sites in Li$_{2}$[(Li$_{0.5}T_{0.5}$)N] for different $T$ elements.}
\label{fig:kmodft}%
\end{figure}

As shown in \rfig{fig:kmmss}, the sign of the MAE contribution from transitions between a pair of subbands
$|m,\sigma\rangle$ and $|m',\sigma'\rangle$ is determined by the spin and orbital character of the involved
orbitals. Because the $d_{z^2}$ orbital is spread out relatively further below the Fermi level and contributes
negligibly to the MAE, we only consider the transitions between subbands with $m$= $-2$, $-1$, 1, and 2. Intra-$|m|$
transitions $|1 \rangle$\lrarrow$|-1 \rangle$ and $|2 \rangle$\lrarrow$|-2 \rangle$ promote easy-axis anisotropy
when they are within the same spin channel, and easy-plane anisotropy when between different spin channels. For
inter-$|m|$ transitions, it is the other way around. Transition $|\pm1 \rangle$\lrarrow$|\pm2 \rangle$ promotes
easy-plane anisotropy when it is within the same spin channel and easy-axis anisotropy when between different spin
channels. The signs and coefficients of the MAE contributions from different orbital pair transitions are indicated
in \rfig{fig:kmodft}(a). Transitions contribute to MAE only when they cross the Fermi level. The amplitude of MAE
depends on the orbital characters and also the energy difference between the two band centers.  When the Fermi
level intersects the narrow degenerate states, the transition energy required to excite an electron across the
Fermi level is very small (between 0 and bandwidth), making the MAE contribution from this pair of orbitals very
large. On the other hand, when the Fermi level is between two well-separated DOS peaks, the required transition
energy is much larger so the amplitude is much smaller.

To elucidate the orbital contributions from the Fe$_{1a}$ site to the MAE in Li$_{2}$[(Li$_{0.5}$Fe$_{0.5}$)N], we
approximate the densities of states of $|\pm1\rangle$ ($d_{xz},d_{yz}$) and $|\pm2\rangle$
($d_{xy}$,$d_{x^{2}-y^{2}}$) subbands with two Lorentzian functions. Crystal-field splitting $\Delta
c$=$\epsilon_{|m|=1}-\epsilon_{|m|=2}$=$1.8$eV, spin splitting $\Delta s$=2.4 eV, and half width $w=0.06$ eV are
used to represent the DFT-calculated PDOS, as shown in \rfig{fig:pdos}. The PDOS used in our model is shown in
\rfig{fig:kmodft}(a) and the MAE contribution from the $1a$ site and its decomposition into orbital pair
transitions as functions of the Fermi energy are shown in \rfig{fig:kmodft}(b). With $T$=Fe, the Fermi level
intersects the $|\pm2,\downarrow\rangle$ states, which results in a large uniaxial anisotropy.  Using \req{eq:k3l},
Fe$_{1a}$ has a MAE contribution which is of the order of 15 $m$eV/Fe.  As shown in \rfig{fig:kmodft}(b), for
$T$= Fe, nearly all MAE contributions are from the transitions
$|2,\downarrow\rangle${\lrarrow}$|-2,\downarrow\rangle$, in other words, between $d_{x^2-y^2}$ and $d_{xy}$
orbitals in the minority spin channel.

To compare with the above analytical modeling, MAE calculations were carried out in both VASP and all-electron
FP-LMTO. The difference of MAE values using two methods is less than $5\%$ for $T$=Fe. To decompose the MAE, we
evaluate the SOC matrix element $\langle V_{so} \rangle$ and its anisotropy $K(\langle V_{so} \rangle)$, which can
be easily decomposed into sites, spins, and orbital pairs\cite{antropov.ssc2014}. We found that $K\approx K(\langle
V_{so} \rangle)/2$ for all $T$ compounds, which suggests that second-order perturbation theory is a good
approximation. As shown in \rtbl{tbl:klitn}, for $T$=Fe, the total MAE is 20.8 $m$eV (per 24-atom cell) and MAE
contributions from $1a$ and $2d$ sites are 14.77 and 3.09$m$eV/Fe, respectively. The contributions from Li
and N atoms are nearly zero as expected. Thus, the impurity Fe (especially Fe$_{1a}$) atoms are essentially the
only MAE providers. By further investigating the matrix element of SOC on the $1a$ site, we found that nearly all
the MAE contributions came from intra-$|m|$ transitions of
{$|2,\downarrow\rangle$\lrarrow$|-2,\downarrow\rangle$}. As shown in \rtbl{tbl:kso}, the $4\chi^{\epsilon}_{22}$
term (dominated by $\chi^{\downarrow\downarrow}_{22}$ for $T$=Fe ) contributes 15.1 $meV$/Fe and the
$\chi^{\epsilon}_{11}$ term has a much smaller negative value of -0.42 $meV$/Fe, while other terms are
negligible. Hence, DFT results agree with our model very well.

With magnetization along the $c$ direction, the SOC can lift the orbital degeneracy and shift two narrow bands
$m=\pm2$, one below and the other above the Fermi level completely, with orbital quantum number $m^c=\pm2$,
respectively, where $m^c$ is the orbital quantum number in the complex spherical harmonics. As a result, the
density of states at the Fermi level becomes very small. Indeed, experiments\cite{jesche.nphy2014} found this
system to be an insulator for $T$=Fe. It had been shown that
\cite{klatyk.prl2002,novak.prb2002,novak.jmmm2004,antropov.prb2014} the correlation effect further enhances the
separation between occupied and unoccupied states. Using the LDA+$U$ method, we also found that correlation can
enhance the orbital moment when the spin is along the axial direction.

Fe concentration and site disordering can significantly affect the MAE. As we have shown, the Fe$_{2d}$ sites,
which represent a high-doping concentration, have much lower anisotropy than the Fe$_{1a}$ sites, which represent a
lower-doping concentration.  By replacing the Fe$_{2d}$ sites back with Li atoms in the $hex2$ supercell, we
calculated the MAE with a smaller concentration $x$=0.166 and found that MAE increase to 22 meV/Fe, which is in
very good agreement with previous calculations.\cite{novak.prb2002}. An interesting concentration is $x$=0.33.  If
only one of two $2d$ sites is occupied by Fe in the $hex2$ supercell, as shown in \rfig{fig:xtal}, then this
configuration would correspond to $x$=0.33 and the supercell has two well-isolated Fe atoms. The DFT calculation
shows high MAE with a value of 20 meV/Fe. On the other hand, if the two Fe atoms occupy the 2d sites and then are
not well separated, the resulting MAE is much smaller (2.8 meV/Fe). Even if we assume that Fe atoms tend to
separate, with a concentration beyond $x$=0.33, it is unavoidable to have Fe atoms neighboring each other and the
hybridization between them causes the MAE (per Fe) to decrease. Furthermore, impurity sites are disordered, as
found in experiments.  At least at a higher concentration, many Fe atoms would not have the symmetric lateral
surroundings as the two Fe sites do in the $hex2$ supercell we used in the calculations. This site disordering may
also have an effect on MAE by lowering the point-group symmetry of Fe impurity sites. And the $m=\pm2$ states on Fe
sites are no longer degenerate, which may decrease MAE per Fe.

\subsection{MAE in Li$_{2}$[(Li$_{1-x}T_{x}$)N] with $T$=Mn, Co and Ni: The band-filling effect}
%% \label{sec:4}

Figure \ref{fig:kmodft}(a) shows how the Fermi level changes with different $T$ elements in a simple rigid-band
picture. Only those transitions across the Fermi level contribute to MAE. With $T$ elements other than Fe, the
$|\pm2,\downarrow\rangle$ states become either fully occupied or unoccupied. The large uniaxial anisotropy that
originated from transition $|2,\downarrow\rangle${\lrarrow}$|-2,\downarrow\rangle$ (term
4$\chi_{22}^{\downarrow\downarrow}$) vanishes and other transitions becomes important, depending on the position of
the Fermi level. For $T$=Ni, the Fermi level intersects the degenerate $|\pm1,\downarrow \rangle$ states.  Hence
anisotropy contributions are dominated by the transitions $|1,\downarrow\rangle${\lrarrow}$|-1,\downarrow\rangle$
(term $\chi_{11}^{\downarrow\downarrow}$). This transition promotes the uniaxial anisotropy, as
4$\chi_{22}^{\downarrow\downarrow}$ does for $T$=Fe.  For $T$ = Co, the Fermi level is between {\pmbd} and
$|\pm1,\downarrow \rangle$ peaks. The transitions of {\pmbd} {\lrarrow} {\pmad} (term
-3$\chi_{12}^{\downarrow\downarrow}$) and {\pmau} {\lrarrow} {\mpad} [term
  -($\chi_{11}^{\uparrow\downarrow}$+$\chi_{11}^{\downarrow\uparrow}$)] support easy-plane anisotropy, while the
transition {\pmbu \lrarrow \pmad} (term 3$\chi_{12}^{\downarrow\uparrow}$) promotes easy-axis anisotropy. However
the two bands involved in the last transition are far away from each other and this contribution is relatively
small. Hence, for $T$=Co, one should expect the system to have easy-plane anisotropy. For $T$=Mn, there are four
transitions that contribute to the MAE; all of them are between the two spin channels, in which two inter-$|m|$
transitions {\pmau \lrarrow \pmbd} (term 3$\chi_{12}^{\uparrow\downarrow}$) and {\pmbu \lrarrow \pmad} support
easy-axis anisotropy, while two other intra-$|m|$ transitions {\pmau \lrarrow \mpad} and {\pmbu \lrarrow \mpbd}
[term -4($\chi_{22}^{\uparrow\downarrow}$+$\chi_{22}^{\downarrow\uparrow}$)] support easy-plane anisotropy. The
four transitions compete and the sign of the total MAE is not obvious and requires a more quantitative description.

The SOC constant $\xi$ changes with element. In \rfig{fig:kmodft}(b), we plot the scaled MAE
$\tilde{K}$=${K}/{4\xi^2}$ and its orbital-resolved components as functions of the Fermi level by using parameters
of $\Delta s$, $\Delta c$, and $w$ for $T$=Fe. In a rigid-band picture, it clearly shows that Ni also has a
uniaxial anisotropy with contributions coming from the $\chi_{11}^{\downarrow\downarrow}$ term.  Since we are using
the same half width $w$ of LDOS for $m$=$\pm1$ and $m$=$\pm2$ subbands, we have $\tilde{K}_{Ni}\approx
\frac{1}{4}\tilde{K}_{Fe}$ because of the intra-$|m|$ transitions coefficients $m^2$, as shown in
Eqs.~(\ref{eqn:amx}) and (\ref{eqn_ke}).  Figure \ref{fig:kmodft}(c) shows the MAE $K$ as a function of the number
of occupied electrons by using different sets of $\Delta s$, $\Delta c$, and $w$ parameters to better present
DFT-calculated PDOS for different $T$ elements, as shown in \rfig{fig:pdos}. The SOC constant $\xi$ is interpolated
by using DFT-calculated $\xi$ values for $3d$ elements. Since $\xi$ decreases with the atomic number within a given
$nl$ shell, ${K}$ quickly decreases with smaller atomic numbers due to the factor $\xi^2$. The DFT MAE values are
also plotted to compare with the modeling MAE function. As shown in \rfig{fig:kmodft}(c), with $T$=Fe parameters,
the modeling MAE (Fe rigid-band approximation) can already correctly describe the MAE trend with different $T$
elements.

Although the RBA predicts the correct easy-axis direction for $T$=Ni, the difference between RBA modeling and DFT
is rather large.  In RBA modeling, $K_{Ni}/K_{Fe}$=$(\xi_{Ni}/\xi_{Fe})^2/4\approx$0.6, while the DFT value
(1.71$meV$/atom) for $T$=Ni is about one order of magnitude smaller than for $T$=Fe. This can be explained as
follows. First, we use the same band width for all DOS peaks in our modeling. In fact, the {\pmad} bands are much
broader than the {\pmbd} bands. The easy-axis anisotropy contribution from the transition between {\pmad} states
decreases with increasing band width. Second, the Ni PDOS deviates from the Fe PDOS more than Mn or Co, so RBA is
less appropriate for $T$=Ni. The spin splitting $\Delta s$ and crystal-field splitting $\Delta c$ are much smaller
in Ni than in Fe. This causes the amplitudes of the negative contributions from {\pmbd \lrarrow \pmad} and {\pmau
  \lrarrow \mpad} to become larger and decrease the total uniaxial anisotropy. As shown in \rfig{fig:kmodft}(c), if
we use a smaller $\Delta s$, smaller $\Delta c$, and larger $w$ to better represent the Ni PDOS calculated from DFT
calculations, then much better agreement between model and DFT values can be reached.

For $T$=Co, the model MAE is about twice the DFT value, probably
because of the simplified model DOS. The orbital-resolved $T_{1a}$ MAE
calculated in DFT are summarized in Table \ref{tbl:kso}. Overall,
there is a qualitative agreement between DFT and the analytical model
for the orbital-resolved MAE values for all $T$ elements. It is
interesting that with $T$=Co, the contribution of the
$4\chi_{22}^{\epsilon}$ term is comparable to that of
$-2\chi_{12}^{\epsilon}$ and $\chi_{11}^{\epsilon}$ in DFT, which is
not expected in the model.  As shown in \rfig{fig:pdos}(c), there is a
small portion of unoccupied {\pmbd} states right above the Fermi level
in the minority spin channel, which makes the
$4\chi_{22}^{\downarrow\downarrow}$ terms comparable to
others. However, this electronic structure detail is not considered in
the simplified DOS we use in modeling. If we neglect the
$4\chi_{22}^{\epsilon}$ terms in DFT, then a better agreement between
modeling and DFT can be achieved for $T$=Co.

Thus, the contributions from well-separated impurity sites with $T$
can be well understood. For $T$=Mn and Co, the easy-plane anisotropy
is a result of competition between different transitions, instead of
being dominated by the intra-$|m|$ transition, which strongly depends
on the bandwidth of the degenerate $|\pm m\rangle$ states that are
intersected by the Fermi level. As a result, the band-narrowing effect
on MAE is not as strong as for $T$=Fe or Ni. As shown in
\rtbl{tbl:klitn}, the contributions from $2d$ sites are comparable or
even larger than $1a$ sites for $T$=Mn and Co.

\begin{table}[ptb]
\caption{Orbital-resolved MAE from the $T_{1a}$ site in Li$_{2}$[(Li$_{0.5}$T$_{0.5}$)N] with $T$=Mn, Fe, Co, and Ni. }%
\label{tbl:kso}
\begin{tabular}
[c]{crclcccc}\hline\hline
&  &  &  & \multicolumn{4}{c}{$K$ ($m$eV)}\\\cline{5-8}%
Term& \multicolumn{3}{c}{Orbital Transition} & Mn & Fe & Co & Ni\\\hline
4$\chi^{\epsilon}_{22}$ & $d_{xy}$ & $\Leftrightarrow$ & $d_{x^{2}-y^{2}}$ &
-0.86 & 15.10 & 0.71 & -0.03\\
$\chi^{\epsilon}_{11}$ & $d_{yz}$ & $\Leftrightarrow$ & $d_{xz}$ & -0.22 &
-0.42 & -0.78 & 3.68\\
-2$\chi^{\epsilon}_{12}$ & $d_{yz}$,$d_{xz}$ & $\Leftrightarrow$ &
$d_{xy}$,$d_{x^{2}-y^{2}}$ & 0.73 & -0.18 & -0.81 & 0.09\\
-3$\chi^{\epsilon}_{01}$ & $d_{z^{2}}$ & $\Leftrightarrow$ & $d_{yz}$,$d_{xz}$ &
0.03 & 0.08 & -0.01 & -0.25\\\hline\hline
\end{tabular}
\end{table}

\bigskip

\section{Summary and Conclusion}
\label{sec:4}

Based on second-order perturbation theory, MAE is resolved into
contributions from different pairs of orbital transitions, more
precisely, the difference between spin-parallel and spin-flip
components of the orbital susceptibilities of the corresponding
orbital pair. In the Li$_{2}$[(Li$_{1-x}T_{x}$)N] systems, with
$T$=Mn, Fe, Co, and Ni, the linear geometry of the $T$ sites minimizes
the in-plane hybridization and results in atomic like orbitals around
the Fermi level for all $T$ elements. The MAE oscillates with the
atomic number from $T$=Mn to $T$=Ni, which is a result of the
competition between contributions from all allowed orbital
transitions. As the Fermi level evolves with $T$, different orbital
pair transitions dominate the contribution to MAE. For $T$=Fe and
$T$=Ni, the intra-$|m|$ transitions within the minority spin channel
dominate the MAE contribution and result in a uniaxial anisotropy. For
$T$=Mn and Co, the easy-plane anisotropy is a result of the
competition between contributions from several transitions with
different signs. Using Lorentzian density of states, we investigate
the band-filling effect on MAE in an analytical model based on a
Green's function technique. We show the MAE as a continuous function
of atomic number. This analytical model can already describe the
correct trend of the MAE obtained using DFT, by just using a simple
rigid Fe band picture. If we take into account the deviation from the
rigid Fe band model and some details of DFT electronic structure, an
even better agreement between the model and DFT can be found. To
further validate our modeling analysis, we also calculate the
orbital-resolved MAE by evaluating the SOC matrix element in
DFT. Overall, Li$_{2}$[(Li$_{1-x}T_{x}$)N], with $T$=Mn, Fe, Co, and
Ni, is a unique system which clearly shows the band-filling effect on
MAE and the nature of this effect can be understood in a very simple
model.

\section*{Acknowledgement}
We would like to thank A. Jesche, P. Canfield, V. Antropov,
A. Chantis, B. Harmon, T. Hoffmann, and D. Johnson for helpful
discussions. Work at Ames Laboratory was supported by the US
Department of Energy, Energy Efficiency and Renewable Energy, Vehicles
Technology Office, Advanced Power Electronics and Electric Motors
program, under Contract No. DE-AC02-07CH11358.  \bibliography{aaa}

\end{document}